# SSBD Ontology: A Two-Tier Approach for Interoperable Bioimaging Metadata

Yuki Yamagata[1,2][0000-0002-9673-1283], Koji Kyoda[3][0000-0001-9455-2153],

Hiroya Itoga[3][0000-0001-5224-3811], Emi Fujisawa[3] and Shuichi Onami[1,3][0000-0002-8255-1724]

[1] Life Science Data Sharing Unit, RIKEN Information R&D and Strategy Headquarters, 2-2-3 Minatojima-minamimachi, Chuo-ku, Kobe, Hyogo 650-0047, Japan
{yuki.yamagata,sonami}@riken.jp
[2] Integrated Bioresource Information Division, RIKEN Bioresource Research Center, 3-1-1 Ko-yadai, Tsukuba, Ibaraki 350-0074, Japan
[3] Laboratory for Developmental Dynamics, RIKEN Center for Biosystems Dynamics Research, 2-2-3 Minatojima-minamimachi, Chuo-ku, Kobe, Hyogo 650-0047, Japan
{kkyoda,hiroya.itoga,emi.fujisawa}@riken.jp

**Abstract.** Advanced bioimaging technologies have enabled the large-scale acquisition of multidimensional data, yet effective metadata management and interoperability remain significant challenges. To address these issues, we propose a new ontology-driven framework for the Systems Science of Biological Dynamics Database (SSBD) that adopts a two-tier architecture. The core layer provides a class-centric structure referencing existing biomedical ontologies, supporting both SSBD:repository—which focuses on rapid dataset publication with minimal metadata—and SSBD:database, which is enhanced with biological and imaging-related annotations. Meanwhile, the instance layer represents actual imaging dataset information as Resource Description Framework individuals that are explicitly linked to the core classes. This layered approach aligns flexible instance data with robust ontological classes, enabling seamless integration and advanced semantic queries. By coupling flexibility with rigor, the SSBD Ontology promotes interoperability, data reuse, and the discovery of novel biological mechanisms. Moreover, our solution aligns with the Recommended Metadata for Biological Images guidelines and fosters compatibility. Ultimately, our approach contributes to establishing a Findable, Accessible, Interoperable, and Reusable data ecosystem within the bioimaging community.

**Keywords:** Ontology, Bioimaging, Metadata.

## 1 Introduction

Recent remarkable advances in bioimaging technologies, such as cryo-electron microscopy and sophisticated live-cell microscopy, have enabled the large-scale observation and recording of biological phenomena at multiple scales (molecular, cellular, tissue-level, and so on). To facilitate global sharing of these vast bioimaging datasets, repositories such as the BioImage Archive (BIA) [1] and Image Data Resource (IDR) [2] have been developed. However, the ontologies referenced in the metadata vary among repositories, and full interoperability has yet to be achieved. Although the community has produced metadata recommendations—such as the Recommended Metadata for Biological Images (REMBI) [3]—a systematic schema or inference mechanism akin to an ontology-based framework is still lacking. In response to these challenges, the foundingGIDE project aims to achieve international harmonization.[1]

We develop and maintain the Systems Science of Biological Dynamics Database (SSBD), a platform that promotes the sharing and reuse of bioimaging data, including microscopy images and image segmentation/tracking data [4]. The SSBD has been reorganized into a two-tier structure:

SSBD:repository - A repository enabling the rapid publication of datasets at the time of paper publication or submission with only minimal metadata.
SSBD:database - A value-added database that offers highly reusable data accompanied by rich ontology annotations.

Through this two-tier structure, the SSBD aims to integrate bioimaging data. In this paper, we present a newly constructed SSBD Ontology, aligned with this two-tier architecture, and propose a framework to enhance and expand existing metadata infrastructure while improving interoperability. Ultimately, our goal is to contribute to the foundation of an imaging data ecosystem under the foundingGIDE project, which seeks international harmonization.

## 2    Issues and Related Research

### 2.1    Domain-Specific Challenges in Bioimaging

Bioimaging inherently involves extremely large-scale images and multidimensional datasets (including spatial, temporal, and channel dimensions). Moreover, describing the relevant experimental methods, genes, reagents, underlying biological structures, and cells associated with each observation requires capturing information from multiple angles. The community faces two simultaneous needs: (1) to rapidly publish data with minimal metadata to align with early-stage research findings, which often emphasize the prompt reporting of cutting-edge techniques, and (2) to add detailed annotations at a later stage to support the elucidation of the biological mechanisms.

In addition, linking with bibliographic author information (PubMed, DOI, etc.) is indispensable. A framework that integrates the various forms of biological information described in this article, including gene expression, organs, cell types, and other observed biological entities, is highly desirable. Although ontology technology is considered a powerful solution for handling the complex and diverse range of information in this domain, it has not yet been fully leveraged in practice. In cutting-edge biological research, in which large-scale imaging data are increasingly combined with genetic and physiological information for multimodal AI analyses, the importance of consistently structured data is expected to grow.

---

[1]   https://founding-gide.eurobioimaging.eu/

### 2.2  Difficulties Associated with the Metadata Template

The SSBD has long relied on a workflow in which annotators manually enter metadata into spreadsheets and the authors confirm the entries. However, this confirmation process often involves the authors providing keyword-based descriptions, leading to natural language variations and inconsistencies. In the latest version of the SSBD metadata[2] (version 3), the scope of the metadata has been expanded to include anatomical structures, genes, and information on the reagents used in experiments, and it complies with REMBI. However, maintaining the metadata solely in a spreadsheet form makes it difficult to manage term consistency, and issues with compatibility between the SSBD and other databases, such as IDR and BIA, remain. To address these issues, we treat the roles of the curator and ontologist as complementary: First, curators (domain biologists) validate the biological accuracy and provide spreadsheet metadata. After review by imaging experts, an ontologist translates these sheets into Resource Description Framework (RDF) using Web Ontology Language (OWL) [5] , building the class hierarchies from the Open Biological and Biomedical Ontology (OBO) Foundry ontologies [6], describing object-property relations, and generating the instance-level assertions described in Section 3.

### 2.3  Related Research

While the SSBD refers to existing biomedical ontologies (e.g., Gene Ontology [7], UBERON [8], Cell Ontology [9], and Cell Line Ontology [10]), the ontology or controlled vocabulary adopted by other databases (e.g., IDR and BIA) differs slightly from SSBD choices. For instance, gap analysis comparing the SSBD with IDR and BIA (REMBI–SSBD mapping) shows that the SSBD uses ChEBI [11] to reference chemical compounds, whereas IDR uses PubChem [12]; thus, a mechanism to bridge such gaps is required.[3] An older version of the SSBD metadata was previously converted into RDF;[4] however, that implementation primarily composed of instance data with minimal or no hierarchical structure, and has not been updated since 2018. It also lacks restrictions on properties, leading to inconsistencies and hindering future extension. Consequently, the legacy version does not accommodate the expanded annotations introduced by the latest REMBI-compliant metadata, which undermines its interoperability in international collaborations. Moreover, the existing ontology fails to clearly distinguish between biological elements and microscopic techniques and has not been regularly maintained to incorporate new resources. These limitations highlight the need for a newly established schema that integrates a robust hierarchical framework, provides domain and range constraints, and can adapt to evolving community standards.

This section highlights the need to link diverse scales and types of related information in the bioimaging domain and to adopt an appropriate approach. We also emphasize building a framework that can bridge gaps among multiple imaging databases and process information in a consistent manner—an aspect that is becoming increasingly important for AI-based analyses.

---

[2] https://github.com/openssbd/ssbd-metadata
[3] https://doi.org/10.5281/zenodo.15553217
[4] https://metadb.riken.jp/

## 3      SSBD Ontology Development

The SSBD Ontology developed in this study is described using OWL 2 axioms and is divided into two layers: a core layer and an instance layer. As of 05/12/2025, the ontology contains 1,478 classes, 522 instances, and 235 properties (116 object properties and 119 data properties).

### 3.1      Designing the Core Ontology

Figure 1 shows an overview of the core structure of the SSBD Ontology. Its key architectural principle is to support a two-tier database, distinguishing between the SSBD:repository for rapid publication and the SSBD:database for high-value-added enrichment.

The SSBD:repository tier facilitates immediate data release by defining a minimal metadata structure around four core classes: project, dataset, paper information, and person. Within this foundational structure, a project is linked to its dataset via the has_dataset_output property and to researchers via the has_project_participant property (a sub-property of RO:0000057) from the Relation Ontology.[5] Each paper information entity is automatically enriched with MeSH Topical Descriptors (retrieved from PubMed), giving curators and users a thematic overview of the associated article. To ensure data integrity, the required cardinalities (e.g., exactly one paper information and at least one dataset per project) are checked by SHACL (Shapes Constraint Language) shapes, while the same constraints remain declared in OWL so that reasoners can perform open-world consistency checking and cross-ontology integration.

Once released, a dataset can be optionally advanced to the SSBD:database tier. This tier enables deep enrichment by linking the core entities to detailed class hierarchies via object properties such as has_biosample_information and has_imaging_method_total_information, and by incorporating controlled vocabularies/ontology terms for biological and imaging descriptors.

Within this tier, information related to bioimaging is divided into "Biological information" (e.g., biosample and reagent) and "Imaging information" (e.g., imaging method, imaging instruments, image dimension, and the storage URI of the dataset). This division enables biology experts to focus on annotating the biological aspects.

The SSBD Ontology is designed to support harmonization efforts across the multiple imaging databases of the foundingGIDE project. To achieve this, it increases interoperability by selectively referencing OBO Foundry ontologies to capture different levels of granularity, such as molecules, cellular components, cells, cell lines, anatomical structures, and organisms. For biosample data, the SSBD Ontology incorporates extensive biological background annotations by primarily reusing terms from the OBO Foundry biomedical ontologies: Gene Ontology (GO), Cell Ontology (CL), UBERON, and NCBI Taxonomy (NCBITAXON) [13]. By reusing dereferenceable identifiers from these resources, the ontology inherently functions as Linked Open Data, enabling direct, instance-level linkage with external knowledge graphs. We demonstrate this capability empirically using a federated query in Section 3.3.

---

[5]      https://github.com/oborel/obo-relations

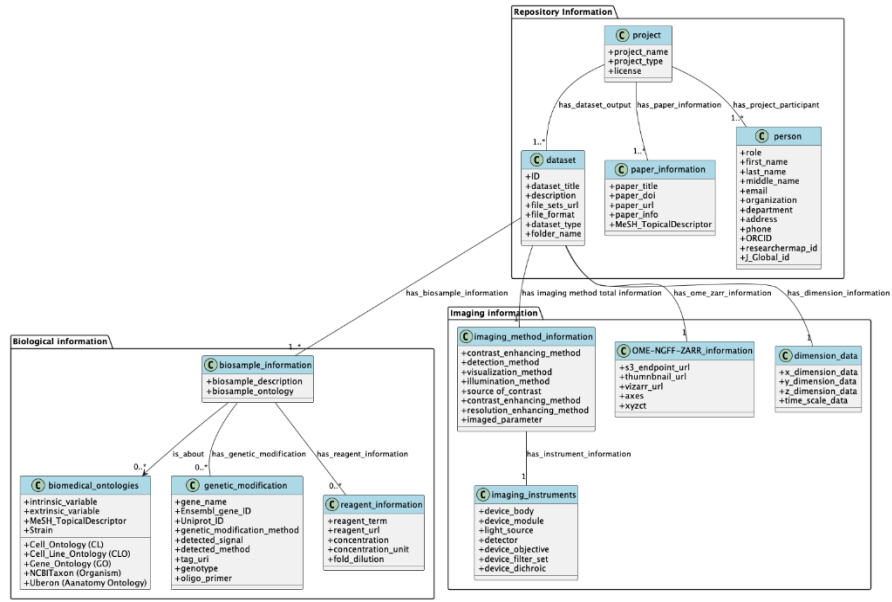

**Fig. 1.** Overview of the core layer of the SSBD

From the perspective of imaging specialists, ontology offers a hierarchical structure for imaging methods, making it possible to categorize large datasets according to the type of microscope or imaging technique used. A key feature of our approach is the integration of semantic metadata with a next-generation file format (OME-NGFF) [14], a cutting-edge cloud-optimized bioimaging format based on Zarr that is capable of efficiently storing massive multidimensional images in object storage [15]. Gaining attention as a Findable, Accessible, Interoperable, and Reusable (FAIR) solution for cloud-based bioimaging data, it serves as a vital link to the SSBD Ontology framework via class SSBD_OME-NGFF-ZARR_information. Each dataset instance, such as dataset-199-Fig2_BrainSliceRGB, is connected (through object properties, such as has_ome_zarr_information) to an OME_NGFF_ZARR individual, which holds key fields, including the S3 object storage endpoint for the OME-Zarr file and a Vizarr viewer [16] link for interactive exploration. As discussed below, SPARQL Protocol and RDF Query Language (SPARQL) queries can traverse the dataset to access its NGFF information, for instance, by retrieving the S3 path.

Accordingly, the constructed ontology can flexibly accommodate both rapid data publication requirements and the rich domain-specific knowledge essential for detailed annotation.

### 3.2 Instance Representation

Metadata sheets describe imaging datasets, which have been converted to RDF instances referencing the core layer. To date, approximately 500 datasets have been processed. Figure 2 provides a concrete example, illustrating how an instance of a project (project-199) is linked to one of its specific datasets (dataset-199-Fig2_BrainSliceeRGB). This dataset instance acts as a hub connecting various metadata entities that includes imaging techniques and other contextual details. For instance, biosample-199 references NCBITaxon:10090 (*Mus musculus*) for taxonomic information. The imaging method is detailed through links to external ontologies, such as FBbi:00000246 for fluorescence microscopy and FBbi:00000304 for the complementary metal oxide semiconductor (CMOS) detection method. Through these connections, each dataset instance specifies the biological subjects involved, the imaging methods, and the associated NGFF storage information required to access the data. This instance layer can thus flexibly represent the unique metadata of each imaging dataset while maintaining structural consistency through the core ontology.

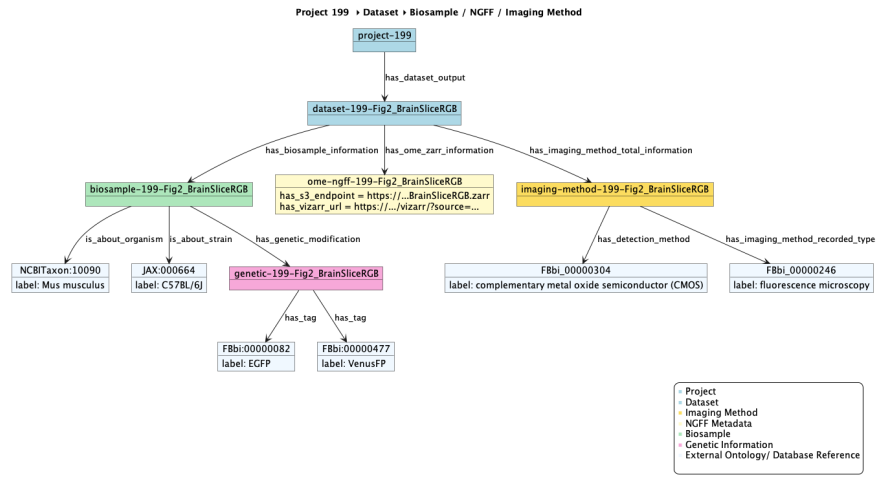

**Fig. 2.** Instance-level view: individuals typed by SSBD core-layer classes and interlinked via object-property assertions

### 3.3 Ontology Evaluation

**Formal Evaluation.** To ensure the formal correctness and consistency of the ontology, we used HermiT [17] to perform OWL 2 reasoning. After several modifications, we confirmed that the ontology is consistent and free of logical contradictions. Specifically, the validation involved checking the class inheritance, domain and range constraints, and data type. One source of error was that some data items did not specify the correct data type (e.g., xsd:anyURI), necessitating minor adjustments to align with the axioms of the ontology.

**Domain Competency Validation.** A critical step in ontology evaluation is to use competency questions to determine whether the ontology satisfies the requirements. To verify how researchers can retrieve bioimaging-related information using the SSBD Ontology, we stored both the core and instance data in an RDF store (Apache Jena Fuseki 4.7.0).[6] We then executed queries using SPARQL to check compliance with these competency requirements. Below are two sample queries and their results, illustrating how the ontology supports both straightforward and complex use cases.

*Query 1: Species-Specific Dataset Search.*[7] As a basic example, this query retrieves all the image datasets in the SSBD derived from *Homo sapiens* samples. The query first identifies all biosamples annotated with NCBITaxon:9606 (*Homo sapiens*) and then selects the datasets associated with them. It proceeds to gather related metadata by traversing up to the parent project and its associated publication. A key feature is the dynamic generation of clickable URLs for any DOI or PubMed ID, allowing direct access to the source literature from the results. Executing this query on the SSBD RDF store returned 12 human-derived datasets, each linked to its project, and providing publication details where available.

*Query 2: Comparing Datasets Across Imaging Modalities.*[8] This more complex case explores how integrated metadata enables the comparison of datasets that use different imaging methods yet share a common strain, such as the C57BL/6J mouse (a common inbred laboratory mouse strain). Using SPARQL, researchers can search for datasets annotated with C57BL/6J but associated with distinct imaging methods. As illustrated in Figure 3, the query results may include one dataset that uses electron microscopy and another that employs fluorescence microscopy. Once the relevant datasets had been identified, direct links to the image data were provided.

Each dataset in the SSBD was associated with a Vizarr viewer URL (see Section 3.1), allowing immediate visualization. Figure 3 couples an imaging-method class hierarchy (right) and a Vizarr snapshot of the real data (left), illustrating how the schema and instance layers together enable semantic-to-visual exploration.

For instance, even if brain tissue images originate from the same anatomical region, researchers may observe different scales, as indicated by the scale bars in the figure. Thus, using the Vizarr links provided in the SPARQL query results, researchers can open the two datasets side by side in a web browser to visually compare them. This use case demonstrates how the SSBD Ontology can address complex queries (e.g., finding datasets associated with different imaging methods) while granting direct access to the underlying image data. As each project record also stores a linked paper resource, users can discover further details about the brain-slice datasets from the trans-scale scope AMATERAS (Project 199-Ichimura [18])—whose publication is recorded as paper-

---

[6] https://jena.apache.org/documentation/fuseki2/
[7] https://github.com/openssbd/ssbd-ontology/blob/main/sparql/organism_human2paper.rq
[8] https://github.com/openssbd/ssbd-ontology/blob/main/sparql/strain2zarr.rq

199-34400683, providing direct access via PMID:34400683 and its DOI—as well as the FIB-SEM electron microscopy datasets from Project 141-Sato [19].

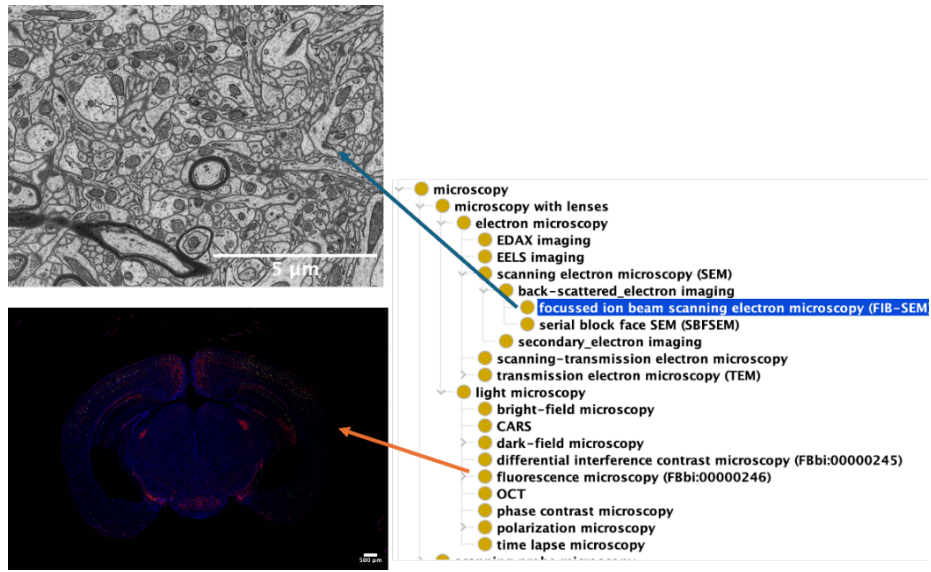

**Fig. 3.** Illustration of the practical application of the SSBD Ontology. This figure combines two distinct elements to demonstrate how the ontology bridges the schema and instance layers: (right) a sub-tree of the imaging-method class hierarchy for detailed annotation, and (left) a Vizarr snapshot of the actual imaging data instance retrieved by the query. This direct connection between abstract semantic descriptions and concrete datasets enables complex, cross-cutting queries. The specific SPARQL query that generates this result and retrieves the Vizarr URL of the dataset is provided in the GitHub repository (sparql/strain2zarr.rq).

Compared with OBI [20], EFO [21], and LOINC [22], the strength of the SSBD is its comprehensive combination of bibliographic data, biological terms, imaging descriptors, NGFF endpoints, and >500 dataset individuals, enabling cross-modal queries such as 'organism → biosample → imaging method → image URI.' In the operational SSBD platform, ontology-anchored metadata are exposed through a public SPARQL endpoint, enabling semantic queries such as "all datasets that have fluorescent tag VenusFP (FBbi:00000477) during calcium-mediated signaling (GO:0019722)". Once these dataset URIs have been retrieved, researchers typically switch to the SSBD REST/NGFF APIs to stream the corresponding OME-Zarr images and BDZarr quantitative tables directly into Python or Napari for downstream statistics and visual analytics. This two-step pattern, SPARQL for discovery and programmatic APIs for bulk access, supports cross-study phenotype screening and lineage-resolved morphometrics.

In future releases, as the ontology is enriched with fine-grained lineage metadata and additional domain-specific terms and entities, this workflow will extend to automated quality control and large language model (LLM) assisted annotations.

In summary, these examples illustrate how the SSBD richly annotates imaging data with high-quality metadata that are structurally organized via an ontology. This consistency supports flexible, domain-relevant queries and provides results that foster new scientific insights.

*Query 3: Cross-ontology Federation – Mitochondrial Biosamples and Related Diseases.*[9] To demonstrate the Linked Open Data capabilities of the SSBD, we executed a federated SPARQL query. This approach is powerful because the SSBD Ontology itself does not contain disease information. Instead, this query connects bioimaging samples to related diseases by leveraging the rich axioms of the MONDO ontology [23]. This is possible because the SPARQL endpoint contains both the SSBD dataset instances and the MONDO ontology, allowing for complex cross-ontology queries within a single data store.

The query operates by first selecting all SSBD biosamples annotated with GO:0005739 (mitochondrion). Then, within the same endpoint, it searches the MONDO ontology for disease classes that are formally defined via RO:0004020 (disease has basis in dysfunction of) as being related to the mitochondria. The result successfully links SSBD biosamples to specific mitochondria-related diseases, such as MONDO:0004069 (inborn mitochondrial metabolism disorder) and MONDO:0044970 (mitochondrial disease). This result illustrates seamless cross-ontology discovery and demonstrates how the SSBD Ontology can contribute to connecting basic bioimaging data with translational disease knowledge.

## 4      Conclusion

In this article, we present the SSBD Ontology supporting the two-tier SSBD structure, comprising the SSBD:repository, which facilitates rapid publication, and the SSBD:database, which offers an added-value resource enriched with comprehensive annotations. We also demonstrate a data validation approach using SPARQL that enables the retrieval of complex life science knowledge associated with bioimaging data. In our implementation, we instantiated approximately 500 datasets, confirming that semantic consistency and FAIR principles can be achieved concurrently for diverse bioimaging data. To broaden discovery and reuse, we will register the ontology with Linked Open Vocabularies (LOV). Future work also includes extension of the datasets, exploring the use of RDF-stars for metadata management, and incorporation of LLMs and Vision Language Models (VLMs) to establish semi-automated annotation pipelines. Our approach is expected to be highly compatible with other databases such as BIA and IDR, thereby contributing to improved data interoperability across the bioimaging community.

---

[9] https://github.com/openssbd/ssbd-ontology/blob/main/sparql/search_mitochondrial_diseases.rq

**Resource Availability Statement:** The SSBD Ontology is available in the NCBO BioPortal repository[10] under the Creative Commons Attribution 4.0 International (CC BY 4.0) license. The ontology and all supporting materials are maintained on GitHub.[11] The repository offers a quick SPARQL walk-through, usage notes, live SPARQL endpoint links, and a CITATION.cff file to facilitate citation. The /sparql/ directory also provides three competency queries (one reproducing Figure 3). A read-only SPARQL service is available online at https://knowledge.brc.riken.jp/bioresource/sparql. The canonical URL for this service is mirrored in the GitHub repository for continuity. For long-term preservation, all ontology resources have been archived in Zenodo (concept DOI: 10.5281/zenodo.15700645[12]), together with the core ontology (classes), instance data, and metadata.ttl (VoID + DCAT). Each release is tagged following semantic versioning and automatically deposited under the same DOI.

**Maintenance:** Supported since 2013, the SSBD has received long-term funding from RIKEN and EU grants; every tagged release is checked with HermiT and archived in Zenodo, guaranteeing reproducibility.

We actively collaborated with the foundingGIDE community for long-term sustainability and a broader impact.

**Acknowledgments.** We express our gratitude to the members of the foundingGIDE project, BioImage Archive, Image Data Resource, and the Open Microscopy Environment for their advice and support. We are also grateful to tools such as DeepL, Grammarly, and ChatGPT, which helped us refine our writing; however, we carefully reviewed and edited their outputs during manuscript preparation. We thank the staff of the Onami Laboratory for their advice and assistance. This study was funded by the National Bioscience Database Center [Grant number: JPMJND2201] and Core Research for Evolutional Science and Technology [Grant numbers: JPMJCR1511 and JPMJCR1926] of Japan Science and Technology Agency; Japan Society for the Promotion of Science [Grant numbers: JP18H05412 and JP22H04926]; and European Union [Grant number: 101130216].

**Disclosure of Interests.** The authors declare no competing interests relevant to the content of this article.

**Author Contributions.** Y.Y.: study design, ontology development, metadata conversion, validation, analysis, and writing; K.K.: metadata modeling, data selection, metadata curation, data conversion, validation, writing, review, and editing; H.I.: system development, data receipt, management and import, review, and editing; E.F.: data selection, metadata curation, and review; S.O.: supervision, writing, review, and editing. All the authors have read and approved the final version of the manuscript.

---

[10] https://bioportal.bioontology.org/ontologies/SSBD
[11] https://github.com/openssbd/ssbd-ontology
[12] https://doi.org/10.5281/zenodo.15700645